# A High-Efficiency Microwave Power Combining System Based on Frequency-Tuning Injection-Locked Magnetrons

Xiaojie Chen, *Graduate Student Member, IEEE*, Bo Yang, *Student Member, IEEE*, Naoki Shinohara, *Senior Member, IEEE*, and Changjun Liu, *Senior Member, IEEE*

*Abstract*—To increase the power level and energy utilization rate of injection-locked magnetron sources, a dual-way 1-kW *S*-band magnetron microwave power combining system with high combining efficiency was proposed and validated. A waveguide magic-Tee was used to achieve power combining and to provide a pathway for the reference signal. This system utilizes the power-dividing characteristic of a magic-Tee to lock two magnetrons. Frequency tuning is applied to adjust the phase difference between the two magnetrons' signals so as to achieve a high combining efficiency. Experimental results indicate that the microwave power combining efficiency of the proposed system reaches 94.5%. The attenuation of microwave power is caused only by the waveguides and magic-Tee. Our investigation provides a guideline for future high-power microwave combining systems with low losses.

## I. INTRODUCTION

CONTINUOUS-WAVE (CW) microwaves have shown their potentials in industrial applications, in which pulsed microwaves are inapplicable or limited, such as acceleration of chemical synthesis reaction rates [1], waste treatments [2], and wireless power transmission (WPT) [3]. To overcome the power limitation of the single microwave source and meet the increasing demand for microwave industrial applications, the power-combining techniques have recently attracted tremendous research attention. Kobayashi *et al.* [4] proposed an *S*-band 1-kW-class solid-state power amplifier

Manuscript received May 14, 2020; revised July 5, 2020 and July 27, 2020; accepted July 29, 2020. Date of publication August 19, 2020; date of current version September 22, 2020. This work was supported in part by the China Scholarship Council under Grant 201906240240, in part by the National Natural Science Foundation of China (NSFC) under Grant 61931009, and in part by the China 973 Program under Grant 2013CB328902. The review of this article was arranged by Editor J. Feng. *(Corresponding authors: Naoki Shinohara; Changjun Liu.)*

Xiaojie Chen is with the School of Electronics and Information Engineering, Sichuan University, Chengdu 610064, China, and also with the Research Institute for Sustainable Humanosphere, Kyoto University, Uji 611-0011, Japan (e-mail: xjchen9112@163.com).

Bo Yang and Naoki Shinohara are with the Research Institute for Sustainable Humanosphere, Kyoto University, Uji 611-0011, Japan (e-mail: yang_bo@rish.kyoto-u.ac.jp; shino@rish.kyoto-u.ac.jp).

Changjun Liu is with the School of Electronics and Information Engineering, Sichuan University, Chengdu 610064, China (e-mail: cjliu@ieee.org).

based on an eight-way power combining which achieved a drain efficiency of 54% and a power combining efficiency of 87%. Traveling-wave tubes (TWTs) and klystrons have been reported to generate CW microwaves of hundreds of kilowatts, but their conversion efficiencies are generally near 40% [5], [6]. The aforementioned microwave devices are more suitable for wideband communication and radar applications than industry, owing to their high cost and low efficiency. In comparison with semiconductors, klystrons, and TWTs, CW magnetrons are economical, high power, and have high dc-to-microwave-conversion efficiency.

The output microwaves of free-running magnetrons are usually noisy. A fascinating method for solving the noisy output of the magnetrons and coherently combining multiple magnetrons under phase control is the injection-locking technology [7]. Treado *et al.* [8] used a 3-dB hybrid ring as a power combiner in a dual-way pulsed magnetron system and achieved a power combining efficiency of 92%. Zhang *et al.* [9] achieved a power combining system via two master–slave locking 2.45-GHz 1-kW magnetrons with a combining efficiency above 92%. Park *et al.* [10] presented that the efficiency of a dual-way power combining system achieved above 93% via two low phase-jitter 2.45-GHz 1-kW injection-locked magnetrons. Liu *et al.* [11], [12] successfully achieved the largest 2.45-GHz CW magnetron system of dual- and four-way power combining thus far. The combining efficiencies were constantly maintained over 90% when the phases and outputs of the magnetrons were tuned and optimized.

In the previous work, high-power ferrite microwave components were connected to magnetrons, such as ferrite circulators [7]–[9], [11], [12]. They were employed to directing the reference signals to lock the corresponding magnetrons. Ferrite phase shifters were utilized to adjust the phases of magnetrons as well [8], [9]. The insertion loss of a ferrite component has been reported to be high and will deteriorate with increased temperature [13], [14]. Either circulators or phase shifters in high-power microwave applications are big, costly, and lossy.

In this article, we propose an approach to magnetron power combining without any large power circulators or phase shifters connected to magnetrons directly. A magic-Tee was employed for dual-way power combining of magnetrons and separating of the reference signal from magnetrons. When locking occurred in both magnetrons, high-efficiency power combining will be obtained by tuning the reference frequency to compensate for the phase difference of the two synchronized signals. The maximum combining and system efficiencies reached 94.5% and 61.9%, respectively.

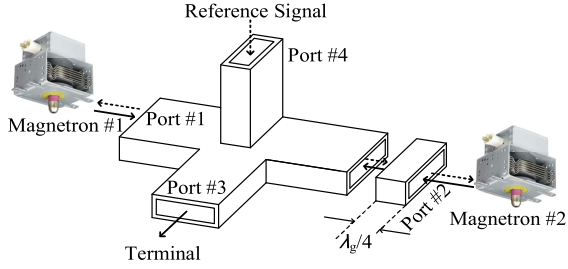

Fig. 1. Schematic of power-combining system based on RMT.

## II. THEORETICAL ANALYSIS

### A. Power Flow of a Magic-Tee

As shown in Fig. 1, a standard magic-Tee with stretching wavelength of $\lambda_g/4$ at Port #2 is employed and named the revised magic-Tee (RMT) for later convenience. We assume that RMT is reciprocal and lossless, then the scattering matrix is defined as follows:

$$\begin{bmatrix} V_1^- \\ V_2^- \\ V_3^- \\ V_4^- \end{bmatrix} = \frac{1}{\sqrt{2}} \begin{bmatrix} 0 & 0 & 1 & 1 \\ 0 & 0 & j & -j \\ 1 & j & 0 & 0 \\ 1 & -j & 0 & 0 \end{bmatrix} \begin{bmatrix} V_1^+ \exp(-j\varphi_1) \\ V_2^+ \exp(-j\varphi_2) \\ 0 \\ V_4^+ \end{bmatrix}$$

$$= \frac{1}{\sqrt{2}} \begin{bmatrix} V_4^+ \\ -jV_4^+ \\ V_1^+ \exp(-j\varphi_1) + jV_2^+ \exp(-j\varphi_2) \\ V_1^+ \exp(-j\varphi_1) - jV_2^+ \exp(-j\varphi_2) \end{bmatrix} \quad (1)$$

where $V_i^+$ and $V_i^-$ are the input and output microwave voltages, respectively, and $\varphi_i$ is the instantaneous phase of the $i$th signal. In Fig. 1, the signal paths are represented by arrows. The dashed arrows are the injected signal paths and the solid arrows are the magnetrons' power paths. The H-arm of the RMT acts as a power combiner. The injected signal fed from Port #4 will be equally divided into two microwave streams to Ports #1 and #2. Ports #3 and #4 are isolated. To obtain the highest combining efficiency, optimal condition should satisfy

$$\frac{1}{\sqrt{2}} V_1^+ \exp(-j\varphi_1) - \frac{j}{\sqrt{2}} V_2^+ \exp(-j\varphi_2) = 0. \quad (2)$$

We assume that $V_1^+ = V_2^+$ and set $\varphi_1$ as reference, then the phase difference $\Delta\varphi$ is deduced

$$\Delta\varphi = \varphi_1 - \varphi_2 = -90°. \quad (3)$$

To achieve the maximum power combining efficiency, $\Delta\varphi$ is supposed to be $-90°$.

### B. Phase Configuration of Injected-Frequency Tuning

We consider the case where a frequency-tunable reference signal is introduced into both magnetrons. When the injected signals reach the magnetrons, the waveform formulas are shown as

$$\begin{cases} V_1(t) = \frac{\sqrt{2}}{2} V_{inj} \cos(2\pi f_{inj} t) \\ V_2(t) = \frac{\sqrt{2}}{2} V_{inj} \cos(2\pi f_{inj} t - 270°) \end{cases} \quad (4)$$

where $V_{inj}$ and $f_{inj}$ are the voltage amplitude and the frequency of the injected signal, respectively.

Here, the frequency of the injected signal is defined as $f_{inj}(t)$, which linearly varies in time and imitates a frequency adjustment [15]. The time-varying frequency of an injected signal is presented as

$$f_{inj}(t) = \begin{cases} f_{inj0}, & 0 < t < t_{start} \\ f_{inj0} + \frac{df_{inj}}{dt}t, & t \geq t_{start} \end{cases} \quad (5)$$

where $f_{inj0}$, $df_{inj}/dt$, and $t_{start}$ are the initial frequency, frequency-change rate, and breakpoint between steady and varying frequencies, respectively. We further obtain the governing equations for the time-varying phase difference and output frequency as

$$\begin{cases} \frac{d\theta_i}{dt} + f_i - f_{inj0} = \frac{\rho_i f_i}{2Q_{iext}} \sin\theta_i, & t < t_{start} \\ \frac{d\theta_i}{dt} + f_i - f_{inj0} - \frac{df_{inj}}{dt}t = \frac{\rho_i f_i}{2Q_{iext}} \sin\theta_i, & t \geq t_{start} \end{cases} \quad (6)$$

$$f_{ioutput}(t) = f_{inj}(t) - \frac{d\theta_i}{dt}. \quad (7)$$

Here, $\rho_i$, $\theta_i$, $f_i$, $f_{ioutput}$, and $Q_{iext}$ are the injection ratio: $(P_{inj}/P_{out})^{1/2}$, the instantaneous phase difference between the $i$th magnetron and the injected frequency, the free-running frequency of $i$th magnetron, the locked frequency of the $i$th magnetron, and the external $Q$ value of the $i$th magnetron, respectively. We can also obtain the well-known Adler's condition [16] when setting $df_{inj}/dt = d\theta_i/dt = 0$. Apparently, the magnetron is locked only when the injected frequency lies within a specific locking bandwidth of $f_i \pm f_i \rho_i/(2Q_{iext})$; when the system returns to a frequency-synchronized state, the phase difference between the injected signal and the magnetron's output is expressed as

$$\begin{cases} \theta_1(t) = 2\pi f_{inj}t - \varphi_1(t) \\ \theta_2(t) = 2\pi f_{inj}t - 270° - [\varphi_2(t) - 90°]. \end{cases} \quad (8)$$

Therefore, the phase difference of the locked magnetrons is

$$\Delta\varphi = \varphi_1 - \varphi_2 = \theta_2 - \theta_1 - 180°. \quad (9)$$

In order to vary the injected frequency to lock two magnetrons, the varying, $df_{inj}/dt$, is supposed to be slower than the response of the frequency synchronization. It is clear that the frequency difference between the reference signal and the magnetron becomes a transient state due to the nonzero $df_{inj}/dt$. To simplify our calculation, we assume that all calculated frequencies are normalized by $f_0$ and the injected strengths $\rho_i/2Q_{iext}$ remain constant. We solved the transient state (6) and (7) to demonstrate the dynamic frequency and phase of each magnetron by the Runge–Kutta method. The initial settings of the reference and free-running magnetron signals are: $f_0 = 1$, $f_{inj0} = 0.9985$, $\rho_i/2Q_{iext} = 0.001$, $f_1 = 0.9995$, and $f_2 = 1.0005$. The injected frequency started to vary at $t_{start} = 2 \times 10^4$. The frequency-change rate in our calculation is set to be $df_{inj}/dt = 1/(6.0 \times 10^7)$. The time-varying frequencies and phase differences of the magnetrons were obtained, as depicted in Fig. 2(a)–(c), respectively.

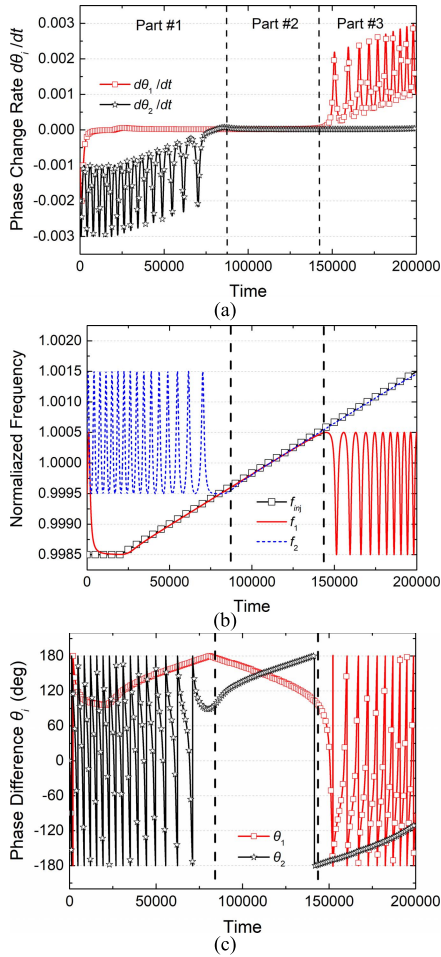

Fig. 2. Numerical predictions of the instantaneous values of (a) $d\theta_i/dt$, (b) $f_{i\text{output}}$, and (c) $\theta_i$ when the injected frequency is linearly varying.

Variation of the phase-change rate $d\theta_i/dt$ can predict the occurrence of phase locking, as shown in Fig. 2(a), such that the output frequency, $f_{i\text{out}}$, can be theoretically plotted in Fig. 2(b). In Fig. 2(c), the wrapped-phase value $\theta_i$ is uniformly distributed between $-180°$ to $180°$ when the magnetrons operate outside the locking bandwidth. When the phase difference $\theta_i$ is synchronized within the locking bandwidth, $\theta_i$ changes regularly; that is, the locking condition was adjusted by linear variation of $\Delta f_i = f_i - f_{\text{inj}}(t)$.

According to the predefined criteria of locking occurrence, the operating process can be divided into three parts. In Part #1, the output frequency of magnetron #1 instantaneously trailed the injected frequency, which varied from 0.9985 to 0.9995. When locking occurred, the phase-change rate $d\theta_1/dt$ equaled zero and the corresponding phase difference $\theta_1$ was dynamically synchronized, owing to the varied $\Delta f$. Conversely, magnetron #2 oscillated at a beat of $\Delta f$ when the injected frequency exceeded the locking bandwidth. Its phase difference, $\theta_2$, fluctuated drastically from $-180°$ to $180°$ due to nonzero $d\theta_2/dt$. Moreover, the locking behaviors of the frequency and phase difference of Magnetron #2 in Part #3 are similar to those of Magnetron #1 in Part #1.

In Part #2, both frequencies $f_1$ and $f_2$ are locked and synchronized with the injected frequency in the overlapped locking bandwidth of 0.9995 to 1.0005. Within the locking bandwidth, we can see that the phase differences $\theta_1$ and $\theta_2$ were simultaneously influenced by variation of $\Delta f$, as shown in Fig. 2(c). Thus, the efficiency of a two-way magnetron power combining system can be adjusted by tuning the injected frequency when the two magnetrons are both locked and tuned.

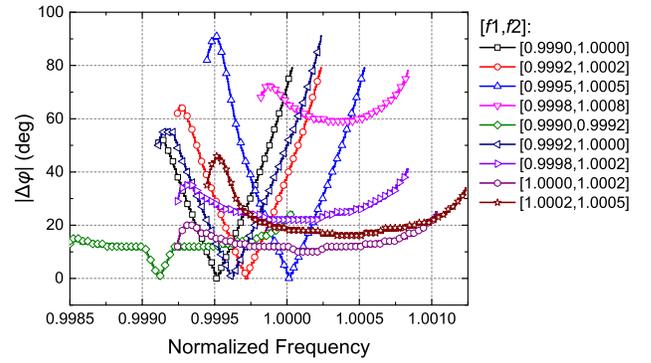

Fig. 3. Numerical iteration of the locking phase difference versus various conditions of initial frequency difference.

The alteration of the phase difference between locked magnetrons with respect to different initial free-running frequencies $[f_1, f_2]$ can be numerically analyzed using (6), where the setup of the injection strengths, $\rho_i/2Q_{i\text{ext}}$, and the frequency variation of the reference signal are the same as in Fig. 2. As shown in Fig. 3, phase-control scopes are obtained due to various initial frequencies and frequencies difference, also indicating that frequency tuning can facilitate valid phase control under unbalanced injection strengths. The power combining efficiency [11] is read as

$$\eta = \frac{1}{2} + \frac{\sqrt{k}\sin|\Delta\varphi|}{1+k} \quad (10)$$

where $k$ is the input power ratio $k = P_1/P_2$. Fig. 3 shows that the frequency tuning can meet the expectation of achieving the high-power combining efficiency when $|\Delta\varphi|$ exceeds 75°. Typically, for the curve of the precondition [0.9990, 0.9992] in Fig. 3, the varied values significantly deviate from expectations for power combining for the RMT. Furthermore, this estimate indicates that an unfavorable phase-control scope might appear in an actual system due to the diversity of unpredictable conditions.

## III. EXPERIMENTAL SYSTEM SETUP

We developed a corresponding experimental system based on WR430 waveguides in a chamber. Fig. 4 shows a block diagram and photograph of the system. The magnetrons (model: 2M167B–M32) were manufactured by Panasonic Microwave Company (Japan) with a 2.45-GHz CW output. The magnetrons were driven using economical switch-mode dc power supplies (WepeX 1000B-TX, Megmeet) with an improved anode-voltage ripple of ∼1%. The filament current can be deactivated after 5-min preheating. A relatively sharp free-running spectrum was achieved. A straight waveguide with a length of 3.75 cm (the wavelength of WR430 waveguide is 14.8 cm at 2.45 GHz) was connected to Port #2 of the standard magic-Tee (5G052, SPC Electronics Company), as shown by the close-up view in Fig. 4(b). Moreover, the characteristics of RMT were measured using a vector network analyzer (VNA) (N9928A, Keysight).

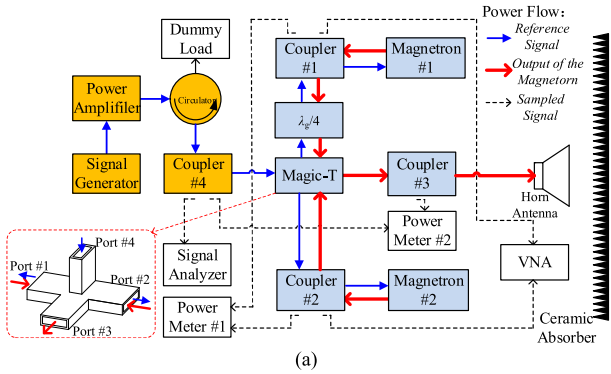

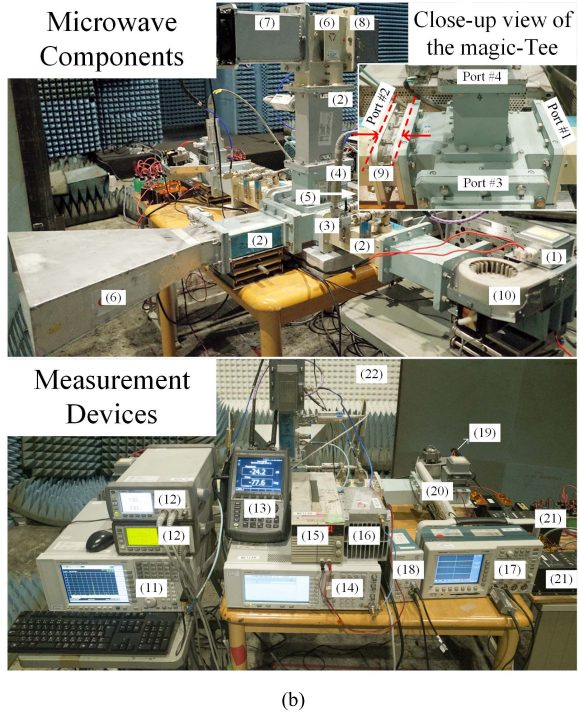

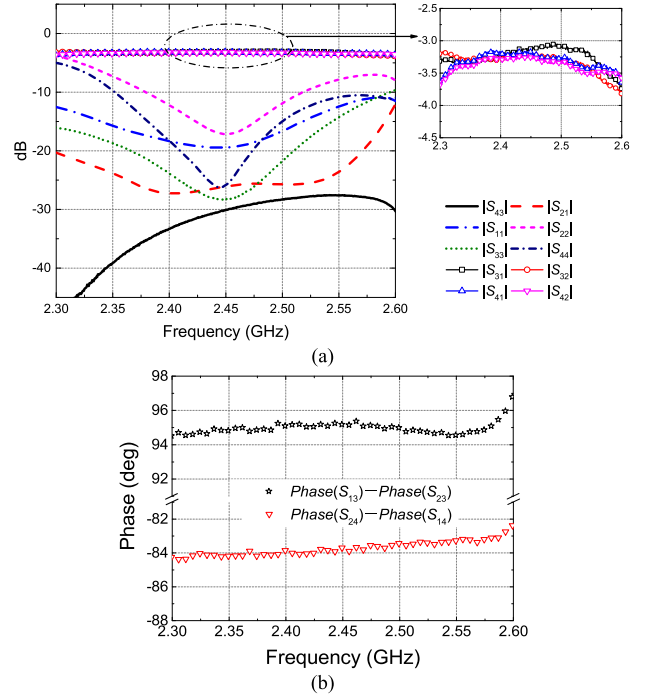

Fig. 4. (a) Block diagram and (b) photograph of the experimental system. Components and devices: (1) magnetron; (2) waveguide coupler; (3) coaxial coupler; (4) power sensor; (5) magic-Tee; (6) horn antenna; (7) wind-cooled dummy load; (8) waveguide to coax adapter; (9) 3.75-cm straight waveguide; (10) fan; (11) signal analyzer; (12) power amplifier; (13) VNA; (14) signal generator; (15) dc power supply; (16) power amplifier; (17) oscilloscope; (18) current-probe amplifier; (19) current probe; (20) high-voltage probe; (21) power supply; and (22) ceramic-absorber wall.

An oscilloscope (TDS-3054, Tektronix) was used to visualize the anode voltage (high-voltage probe: P6015A, Tektronix) and current (ac/dc current probe: TCP312A and current amplifier: TPC A300, Tektronix). A reference signal was produced using a signal generator (E4421B, Agilent) and amplified using a power amplifier (CA2450BW100-4547M-C, R&K). The circulator was connected to Port #4 of RMT to protect the solid-state amplifier and was, therefore, not connected to the magnetrons. Couplers were used to sample the signals and to measure the power and spectrum using power meters (A1914A, Agilent and E4419B, Agilent) and a signal analyzer (N9010A, Agilent), respectively. The horn antenna was used to radiate the combined power. The radiated power was absorbed by the high-power ceramic-absorber wall. Simultaneously, the phase difference of the magnetrons' output was measured using a VNA (N9928A, Keysight).

Fig. 5. Measured results of RMT. (a) Scattering parameters of RMT. (b) Phase difference between the ports of H-arm and the ports of E-arm.

## IV. Experimental Results and Discussion

The measured characteristics of the used magic-Tee are shown in Fig. 5(a) and (b). At the frequency band close to 2.45 GHz, the return losses ($|S_{11}|$ and $|S_{22}|$) of the input ports (Ports #1 and #2) were better than 15.0 dB, and those ($|S_{33}|$ and $|S_{44}|$) of the combined-output (Port #3) and injection (Port #4) ports were better than 23.0 dB. Moreover, both the isolations ($|S_{21}|$ and $|S_{43}|$) were better than 25.0 dB. The transmission coefficients ($|S_{31}|$ and $|S_{32}|$) of the combining branch were approximately $-3.2$ dB and the insertion loss was around 0.2 dB. The measured phase differences between input signals show that the optimal combining efficiency can be obtained when Phase($S_{13}$)− Phase($S_{23}$) was approximately 95.0°. The measured phase difference between the divided reference signals, that is, Phase($S_{24}$)− Phase($S_{14}$), was approximately −84.0°.

The measured spectra are shown in Fig. 6. The spectrum of each magnetron was recorded, while the other one was deactivated. In Fig. 6(a), it is clear that the two magnetrons operated at different frequencies under free-running conditions. The magnetrons, respectively, generated 650- and 665-W CW microwave power, and their initial frequency difference was 1.94 MHz. Significantly, the free-running output spectra were noisy with a visible frequency chirp at around the oscillation frequencies. When two magnetrons were both active in the absence of a reference signal, intermodulation products appeared due to the lack of perfect isolation between Ports #1 and #2, as plotted in the upper-left region in Fig. 6(a). The combining power stability under this condition changed drastically for 5-min monitoring using the LabVIEW-controlled power meter. The raw data detected from the power meter are presented in Fig. 7. The combining efficiency was absolutely unstable and the peak-to-peak value of the power jitters reached 279 W. Furthermore, the unexpected intermodulation significantly increased the risk of electromagnetic interference with the other electrical devices operating in adjacent frequency bands.

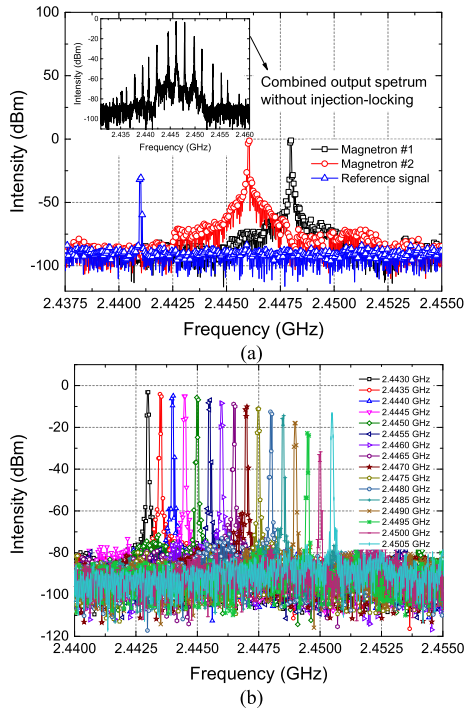

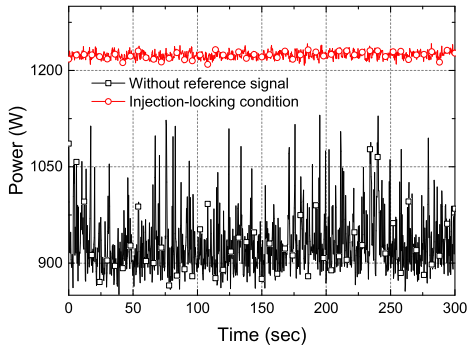

Fig. 6. Measured spectra under different conditions. (a) Spectra of the reference signal, each free-running magnetron, and the condition in which the two magnetrons were both active. (b) Injection-locked spectra of the combined signal during frequency tuning.

Fig. 7. Stability of the combined power.

The reference power fed into Port #4 in our experiments was maintained at 26 W. Unlike the work in [9], where phase adjustment produced intermodulation products, Fig. 6(b) clearly shows that the spectra of the combined outputs remained sharp and clean during reference frequency tuning. When locking occurred, all frequency chirps and inconsistencies between the two magnetrons disappeared. The frequency of the reference signal was tuned with an interval of 500 kHz at constant power. It is noteworthy that the spectra of the experiments were sampled from the reflection signal of Port #4 of the magic-Tee; thus, the reflection amplitude decreased while the combined output increased. At a locking frequency of 2.45 GHz, the peak value of the reflection spectrum approximately equaled the reference spectrum in Fig. 6(a), indicating that the optimal combining efficiency was obtained. The optimal combining power reached 1.230 W with a combining efficiency of 93.5%; moreover, the power stability is shown in Fig. 7. The output-power level remained in a stable condition for 5 min with a peak-to-peak value of alteration of 33 W.

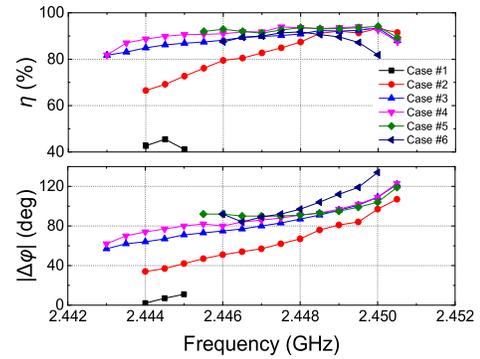

Fig. 8. Variation of combined efficiency $\eta$ and phase difference $|\varphi|$ between injection-locked magnetrons #1 and #2.

Table I shows the experimental results of microwave power combining with respect to different initial power levels. Compared with the dual-way magnetron power combining system developed in [8], [9], [11], and [12], the insertion losses of the circulators or phase shifters were avoided. Under the same injection condition, the high-efficiency power combining was invalid in Case #1, whereas the frequency-tuning and phase-adjustment scopes were narrow and incapable of covering the required phase differences for high-efficiency power combining. The maximum output power and combining efficiency reached 1.800 W and 94.5%, respectively. Loss of the power combining efficiency was mainly caused by the insertion loss of the H-arm of the RMT. The maximum dc-to-microwave efficiency reached 61.9% in Case #4. In Case #6, the microwave output power increased along with the driving power; the power-combining efficiency was still 91.8%, but the dc-to-microwave efficiency decreased to 57.1% due to the decrease in the dc-to-microwave-conversion efficiencies of the magnetrons. It is noteworthy that the power combining efficiency of Case #5 was 1% higher than that of Case #3 even if the power ratio of Case #5 ($k = 1.32$) is higher than that of Case #3 ($k = 1.0$). It can be theoretically explained using (10), when $k = 1.32$, $\Delta\varphi = 104°$, the combining efficiency is 0.77% higher than the condition of $k = 1.0$, and $\Delta\varphi = 109°$. The measured data were similar to the theoretical prediction.

The measured phase difference between the locking magnetrons and the corresponding combining efficiency based on frequency tuning is shown in Fig. 8. The initial-frequency differences between $i$th free-running magnetron and goal reference frequency $f_{\text{goal}}$ is listed in Table I. Case #2 presents a maximum valid phase-control scope of 73°; however, the overlapped locking bandwidth was narrow in Case #1, resulting in a small phase-control scope that failed to compensate $\Delta\varphi$ to the expected value. Increasing the power of the reference signal might effectively solve the problem of insufficient locking scope, but it would significantly increase the cost of the subsystem. It is interesting to note that the optimal combining efficiencies of the different cases appeared at $95° < \Delta\varphi < 110°$, which deviated from the optimal $\Delta\varphi$ by around 10°. This deviation was mainly induced by mismatch of the waveguide connections. Here, we experimentally proved that phase controls of two magnetrons were valid and simultaneous in our proposed system.

TABLE I
MEASUREMENTS OF PROPERTIES OF DIFFERENT CASES

| Case # | Injected Power (W) | Optimal Locking Frequency $f_{goal}$(GHz) | Deviation between free-running frequencies $f_{goal} - f_{ifree-running}$(MHz) | | Magnetron Output Power (W) | | [1] Driven Power (W) | | Combined Microwave Power (W) | Microwave Combined Efficiency (%) | DC to Combined Microwave Efficiency (%) |
|---|---|---|---|---|---|---|---|---|---|---|---|
| 1 |  | 2.4445 | −0.36 | −3.50 | 415 | 640 | 722 | 965 | 480 | 45.5 | 26.9 |
| 2 |  | 2.4500 | 5.64 | 2.00 | 535 | 665 | 790 | 1014 | 1120 | 93.3 | 60.6 |
| 3 | 26 | 2.4500 | 3.94 | 2.00 | 650 | 665 | 977 | 1014 | 1230 | 93.5 | 60.5 |
| 4 |  | 2.4495 | 1.46 | 1.95 | 740 | 665 | 1077 | 1014 | 1320 | 93.9 | 61.9 |
| 5 |  | 2.4500 | 2.24 | 2.00 | 880 | 665 | 1350 | 1014 | 1460 | 94.5 | 60.7 |
| 6 |  | 2.4480 | 0.02 | −2.56 | 980 | 980 | 1580 | 1527 | 1800 | 91.8 | 57.1 |

[1]The driven power was calculated by using measured anode voltage and current of the magnetrons.

The behaviors of the established frequency-tuning power combining system, such as the frequency-pushing effect, unbalanced quality factors, and stability of oscillation, are complicated to compute. However, our measurements of variations of the combined efficiencies and phase differences with respect to frequency tuning were qualitatively similar to the numerical predictions. Importantly, the only major challenge facing the extension of such systems to CW microwave power combining at power levels as high as dozens of kilowatts or megawatts (as presented in [8], [11], and [12]) is the insertion loss of the magic-Tee waveguide, as well as designing the magic-Tee with an artificial cooling system (wind or liquid cooling).

## V. CONCLUSION

In this article, a dual-way 1-kW S-band magnetron power-combining system was built and measured. A RMT waveguide was employed for power combining of the two magnetrons while simultaneously providing paths for the reference signals. The system's behaviors were numerically extrapolated. A stable combined output power and pure spectra during phase control were achieved via frequency tuning of the reference signal. The proposed magnetron coherent power combining system exhibited a maximum microwave combining efficiency of 94.5% and a maximum phase-control scope of 73°, respectively. The power attenuation of each magnetron was minimized in the absence of circulators and phase shifters.

Furthermore, the proposed power-combining system also provided an approach for further establishing low-cost and high-power level microwave sources using industrial magnetrons and magic-Tee to meet the increasing power expectations of the microwave industry.